\documentclass[sigplan]{acmart}

\usepackage[utf8]{inputenc}
\usepackage{xspace}
\usepackage{float}
\usepackage{url}
\usepackage{alltt}
\usepackage{subcaption}
\usepackage{listings}
\usepackage{algorithm}
\usepackage{stmaryrd}
\usepackage{enumitem}
\usepackage{colortbl}
\usepackage{multirow}

\definecolor{amber}{rgb}{1.0, 0.75, 0.0}
\definecolor{almond}{rgb}{0.94, 0.87, 0.8}
\usepackage{xcolor, soul}
\sethlcolor{almond}

\newcommand{\eg}{\emph{e.g.},\xspace}
\newcommand{\ie}{\emph{i.e.},\xspace}
\newcommand{\wrt}{\emph{w.r.t.}\xspace}
\newcommand{\cf}{\emph{cf.}\xspace}
\newcommand{\aka}{\emph{a.k.a.},\xspace}
\newcommand{\etal}{\emph{et.al.},\xspace}

\usepackage{xcolor}

\definecolor{codegreen}{rgb}{0,0.6,0}
\definecolor{codegray}{rgb}{0.5,0.5,0.5}
\definecolor{codepurple}{rgb}{0.58,0,0.82}
\definecolor{backcolour}{rgb}{0.95,0.95,0.92}

\lstdefinestyle{mystyle}{
    commentstyle=\color{codegreen},
    keywordstyle=\color{magenta},
    numberstyle=\tiny\color{codegray},
    stringstyle=\color{codepurple},
    basicstyle=\ttfamily\scriptsize,
    breakatwhitespace=false,         
    breaklines=true,                 
    captionpos=b,                    
    keepspaces=true,                 
    numbers=left,                    
    numbersep=5pt,                  
    showspaces=false,                
    showstringspaces=false,
    showtabs=false,                  
    tabsize=2,
    frame=single
}

\newcommand{\xdeux}{\texttt{x264}\xspace}

\definecolor{fblue}{rgb}{0.0, 0.45, 0.73}
\usepackage{bm}

\usepackage[disable]{todonotes}

\usepackage{xargs}

\newcommandx{\other}[2][1=]{\todo[author=PC,inline, linecolor=blue!40,backgroundcolor=blue!40,bordercolor=blue!40,#1]{#2}}
\newcommandx{\revFour}[2][1=]{\todo[author=R4,inline, linecolor=cyan!40,backgroundcolor=cyan!40,bordercolor=cyan!40,#1]{#2}}
\newcommandx{\xht}[2][1=]{\todo[author=XHT,inline, linecolor=yellow,backgroundcolor=yellow!70,bordercolor=yellow,#1]{#2}}
\newcommandx{\revOne}[2][1=]{\todo[author=R1,inline, linecolor=purple!50,backgroundcolor=purple!50,bordercolor=purple!50,#1]{#2}}
\newcommandx{\revTwo}[2][1=]{\todo[author=R2,inline, linecolor=red!50,backgroundcolor=red!50,bordercolor=red!50,#1]{#2}}
\newcommandx{\revThree}[2][1=]{\todo[author=R3,inline, linecolor=orange!50,backgroundcolor=orange!50,bordercolor=orange!50,#1]{#2}}

\usepackage{algorithm} 
\usepackage{algpseudocode} 

\algnewcommand\algorithmiccreate{\textbf{Create:}}
\algnewcommand\CREATE{\item[\algorithmiccreate]}
\algnewcommand\algorithmicinput{\textbf{Input:}}
\algnewcommand\INPUT{\item[\algorithmicinput]}
\algnewcommand\algorithmicoutput{\textbf{Output:}}
\algnewcommand\OUTPUT{\item[\algorithmicoutput]}
\algnewcommand\algorithmicrepeatt{\textbf{Repeat:}}
\algnewcommand\REPEAT{\item[\algorithmicrepeatt]}
\algnewcommand\algorithmicuntill{\textbf{Until:}}
\algnewcommand\UNTIL{\item[\algorithmicuntill]}

\usepackage{rotating}
\usepackage{diagbox}
\usepackage{graphicx}
\newcommand*\rot{\rotatebox{90}}

\usepackage{amsmath}
\definecolor{ao}{rgb}{0.0, 0.5, 0.0}
\hypersetup{
    colorlinks = true,
    linkbordercolor = {white},
    linkcolor = blue,
    urlcolor = blue,
}

\AtBeginDocument{%
  }

\setcopyright{acmcopyright}
\copyrightyear{2018}
\acmYear{2018}
\acmDOI{XXXXXXX.XXXXXXX}

\acmConference[Conference acronym 'XX]{Make sure to enter the correct
  conference title from your rights confirmation emai}{June 03--05,
  2018}{Woodstock, NY}
\acmPrice{15.00}
\acmISBN{978-1-4503-XXXX-X/18/06}

\setcopyright{none}
\renewcommand\footnotetextcopyrightpermission[1]{} 
\begin{document}

\title{Specialization of Run-time Configuration Space at Compile-time: An Exploratory Study}

\author{Xhevahire Tërnava}
\affiliation{%
  \institution{Univ Rennes, CNRS, Inria, IRISA}
  \city{F-35000 Rennes}
  \country{France}}
\email{xhevahire.ternava@irisa.fr}

\author{Mathieu Acher}
\affiliation{%
  \institution{Univ Rennes, CNRS, Inria, IRISA\\Institut Universitaire de France (IUF)}
  \city{F-35000 Rennes}
  \country{France}}
\email{mathieu.acher@irisa.fr}

\author{Benoit Combemale}
\affiliation{%
  \institution{Univ Rennes, CNRS, Inria, IRISA}
  \city{F-35000 Rennes}
  \country{France}}
\email{benoit.combemale@irisa.fr}


\begin{abstract}
Numerous software systems are highly configurable through run-time options, such as command-line parameters. 
Users can tune some of the options to meet various functional and non-functional requirements such as footprint, security, or execution time.   
However, some options are never set for a given system instance, and their values remain the same whatever the use cases of the system. 
Herein, 
we design a controlled experiment in which 
the system's run-time configuration space can be specialized at compile-time and combinations of options can be removed on demand. 
We perform an in-depth study of the well-known \xdeux video encoder and quantify the effects of its specialization 
to its non-functional properties, namely 
on binary size, attack surface, and performance 
while ensuring its validity.
Our exploratory study suggests that the configurable specialization of a system has statistically significant benefits  
on most of its analysed non-functional properties, which benefits depend on the number of the debloated options.
While our empirical results and insights 
show the 
importance of removing code related to unused run-time options to improve software systems, an open challenge is to further automate the specialization process.
\end{abstract}




\settopmatter{printfolios=true} 
\maketitle

\section{Introduction}
\label{introduction}

Modern software systems are highly configurable and expose to the users their abundant configuration options. 
By setting configuration options' values, a software system is customized for different users to reach specific functionality or performance goals (\eg execution time, energy consumption, quality of the result) without the need to modify its source code.  
Such systems have a large and different number of options.
A considerable number of their options are set at run-time, for example, 
a recent version of \texttt{curl} has about $205$ compile-time options and $242$ run-time options~\footnote{Documentation of curl: \url{https://curl.se/}.}, then the \xdeux video encoder has $39$ compile-time options and $162$ run-time options~\footnote{\url{http://www.chaneru.com/Roku/HLS/X264_Settings.htm}.}.
\revThree{l. 72f It would be interesting to know, how many (run-time + compile-time + ..) options these tool have in total. Are most of them run-time options?}
Run-time options are typically accessible via command-line parameters, configuration files, or menu preferences. 
Such run-time variability is a strength since system's users have the flexibility to tune their systems, owing to their specific use case and deployment constraints.

However, \emph{(i)} there is evidence that "\textit{a significant percentage (up to $54.1\%$) of parameters are rarely set by any use}r"~\cite{xu2015hey} and thus they unnecessarily "bloat" the software. 
\revTwo{The overall motivation seems to be "there are a lot of options and they
are rarely used". I find this motivation somewhat problematic. On the one hand,
yes, great, it gives context to which options may or may not be worth
keeping/maintaining. Though, from evolving programming languages, and perhaps
more specifically APIs, we know that "every feature will be used by someone".
Thus, that there's not direct openly accessible evidence that something is
used, is a very weak signal. Any deprecation, whether language feature or API,
or even the change of internal, undocumented, unspecified behavior will break
some users. Thus, for widely adopted systems it's never "oh this isn't used",
but rather a cost-benefit calculation. Do we gain enough reduction in
maintenance reduction, performance improvement, or similar to justify breaking some users?}
Some options are never tuned and their values remain the same for a given usage, 
for example, depending on the user, the performed activities, and the targeted environment. 
For instance, a user or an organization may choose always to encode a video with \xdeux in an (ultra-)fast way. 
In this case, 
its run-time configuration option of \mbox{-}\mbox{-}\texttt{cabac} is always disabled, as it offers better compression, but requires more processing power to encode and decode a video.
%
Apart from that, \emph{(ii)} it is common to think that a library or a configuration option that is unused today can be used in the future. In fact, quite the opposite is true. 
Pieces of evidence show that the libraries that are unused today in a software system (86\% of them) are unlikely to be ever used in the future~\cite{soto2021comprehensive}.
Then, the unused functionalities in a software system can threaten its security, slow down its performance, affect its reliability, or increase its maintenance~\cite{qian2020slimium}. 
Thus, run-time variability is sometimes unnecessary and does not have to be embedded.
 
Based on this observation, 
we bring up the idea of \emph{specializing the run-time configuration space} of a software system. The goal is to retain only a subset of configurations that meet a functional or a performance requirement and thus discard the rest. Specifically, we aim to debloat run-time options that are never used within the source code, at compile-time.
\revThree{The paper, in most parts, describes the advantages of compile-time variability over run-time variability. However, there could be more room for describing the drawbacks, e.g., managing and distributing the variants of the compiled tools and the need for recompiling the tool to choose an option that was not intended/foreseen to be chosen}
 Which options are unused depend on the system's usage context and they are inputs to the specialization process.
Our hypothesis is that the code 
of unused run-time options is a manifestation of code bloat that may increase the binary size, 
the attack surface, or slow down the system. 
\revOne{I did not fully understand how the quality criteria for evaluation have been chosen? Why is attack surface an important consideration for a video encoder?}
For instance, options that are never used induce dead code that can be eliminated (\eg by a compiler), thus improving the overall system.
This paper aims to 
investigate this hypothesis and explore to what extent removing run-time options can bring benefits to the non-functional properties of software. To the best of our knowledge, quantifying the benefits (if any) has received little attention.

Specializing the configuration space of a software system has a long tradition since the seminal paper of Czarnecki~\etal~\cite{czarnecki2005,czarnecki-helsen-etal2005} and others (\eg~\cite{hubaux2011,amand:hal-01990767,acher2018,DBLP:conf/splc/HarmanJKLPZ14,temple2016,capilla2011promise,temple2017a,martin2021}). 
However, most of the works focused on the specialization of variability models where constraints among individual options (or across several options) are added to enforce the configuration space. 
A missing part we investigate in this paper is to propagate this specialization to the source code of a configurable system. 
Specialization can also be seen as a debloating problem where a subset of run-time options to specialize are bloat of a configurable system. 
Several debloating approaches are proposed in the literature.
Most of them provide a way for debloating the unused functionalities from the external libraries or from high-level "features" of a system~\cite{jiang2018reddroid,sharif2018trimmer,quach2019bloat}. 
Yet, to the best of our knowledge, debloating run-time options of the configurable systems has received little attention. 
An important specificity of our problem is that the specialization should be flexible and is itself configurable. That is, not all options are set once and for all: in contrast, existing debloating techniques specialize the system under study with a full configuration. 
\revOne{On page 2, the authors differentiate their proposal from existing work like so: "That is, not all options are set once and for all: in contrast, existing debloating techniques specialize the system under study with a full configuration." I don't understand how this is a significant and substantive difference? What changes substantively when only a subset of the existing options is considered?}
Our proposal is to annotate run-time options with compilation directives throughout the source code. 
As such, the configuration space of run-time options can be specialized at compile-time and (combinations of) options can be removed on-demand. In the future, the support for automation of these processes asks for software language engineering (SLE) techniques.
 
To realize the idea of specializing a 
configuration space, several new challenges need to be addressed, such as \textit{(i)}~to locate the run-time options within the source code and \textit{(ii)}~to take care of the system's validity after its specialization. 
Our research methodology is to statically annotate run-time options to mitigate the risk of synthesizing an unsound and incomplete specialization. 
In this way, this controlled effort limits the introduction of errors that could bias the benefits (if any) of specialization on non-functional properties. 
In addition, we establish a ground truth for future automatic program specializers and we report on insights from our experience on the case of \xdeux. 

The contributions of this paper are as follows:
\begin{itemize}
    \item We propose
    a means for configuring the specialization of a software system through debloating its unused run-time configuration options at compile-time;
    \item We analyze the resulting non-functional properties (\ie binary size, attack surface, and performance)
    under two specialization scenarios of \xdeux, showing improvements of the \xdeux configurable system without sacrificing its interoperability and validity;
    \item We made available the data for reproducing the experiments and call to replicate our study for further confirming or refuting our results~\footnote{Companion page: \url{https://anonymous.4open.science/r/x264-6ED7}}.
\end{itemize}


\section{Background and motivating case study}
\label{overview}
In this section, we introduce the basic concepts of configurable software systems and the system used as a case study. 

\revTwo{Throughout reading the paper, I was also wondering e.g. "but isn't this what
software product lines do" and "how does this all relate to slicing etc". The
paper may benefit from moving some of the related work techniques to the
background section so that the reader may know early on how these techniques
relate to the proposed work.}

\subsection{Background}
\label{background}
Users can configure a software system by setting the values of its numerous options.
\revThree{It is not entirely clear what the assumptions on the run-time options to apply the proposed procedure are. \\
- What about cross-influences between options? (e.g., --foo can only be chosen if --bar is chosen, too) \\
- What about "non-binary" options, such as, --low, --medium, --high, where only one can be chosen? \\
- What about non-Boolean options of a program (e.g., -size 64)? Sec. 2.1 does not exclude them but the example only contains Boolean options and the algorithm is limited to Boolean options, too.}
Usually, an option carries a particular functionality, has a type of value (namely, Boolean, integer, or string), and a binding time (namely, compile-time or run-time).
A \emph{configuration} is an assignment of values to options at their specific binding time. Default values are also assigned to some options.
Hence, users can build their custom software by (de-)activating some options at compile-time or run-time, but not only at these times~\cite{bosch2012dynamic}. 
The interest of compile-time options is to embed in the resulting executable program what is necessary for a given use case. Typically, preprocessor directives (\eg \#ifdef in C/C++ projects) are used to implement them. 
As for the run-time options, they are (de-)activated prior to, or even during, the execution. Command-line parameters, plugins, or configuration files are examples of mechanisms that users can rely on to control them. 
In this work, we considered the run-time 
options, which are set as command-line parameters during the software's load-time.
Within the source code, run-time options are implemented by ordinary control statements, such as \textit{if} statements, and locating them in code can be challenging as they are implicit.
Nowadays software systems, such as \xdeux video encoder, 
are configurable and often provide compile-time and run-time options to customize them.


\subsection{Motivating case study}
\label{casestudy}

As a motivating system for this study is chosen the software system of \xdeux~\footnote{\url{http://www.chaneru.com/Roku/HLS/X264_Settings.htm}}. It is a command-line video encoder implemented in \texttt{C}, which has been used for the past 5 years to evaluate the debloating approaches~\cite{debloating2021}, is studied among the highly configurable systems (\eg~\cite{guo2013,siegmund2015,temple2017a,jamshidi2017transfer,alves2020sampling,lesoil2021deep}), and has plenty of run-time options. 
We use the \xdeux at its recent commit 
\hyperlink{https://github.com/mirror/x264}{\color{fblue}\underline{\texttt{db0d417}}}~\footnote{\xdeux software system: \url{https://github.com/mirror/x264}}. It has 114,475 lines of code (LoC)~\footnote{Measured using the \texttt{clocl} tool: \url{https://github.com/AlDanial/cloc}}, 237 files, 39 compile-time options, and 162 run-time options.
For example, it is possible to deactivate the support of \texttt{mp4} format at compile-time by using the \mbox{-}\mbox{-}\texttt{disable}\mbox{-}\texttt{lsmash} option during the system build, as in the following.
\begin{lstlisting}[label=listingA, caption={}, language=C, numbers=none]
$ ./configure --disable-lsmash && make
\end{lstlisting}
After the system's build, the run-time options can be set. 
They have an effect on the properties of the encoded video, namely on its encoding time, bitrate, and frame rate.
Here is a possible usage of three run-time options in \xdeux.
%
\begin{lstlisting}[label=listingB, caption={}, language=C, numbers=none]
$ x264 --cabac --mbtree --mixed-refs -o vid.264 vid.y4m
\end{lstlisting}
%
The used \mbox{-}\mbox{-}\texttt{cabac}, \mbox{-}\mbox{-}\texttt{mbtree}, and \mbox{-}\mbox{-}\texttt{mixed}\mbox{-}\texttt{refs} options are Boolean. They also have their 
variant to negate their functionality, for example, \mbox{-}\mbox{-}\texttt{mbtree} has \mbox{-}\mbox{-}\texttt{no}\mbox{-}\texttt{mbtree}.
\revThree{l. 230ff: Why is it important to mention that the Boolean options have negated variants? Why are the negated variants needed? E.g., what is the difference between choosing --no-mbtree and choosing neither --mbtree nor --no-mbtree?}

\xdeux is a free software application for encoding video streams into the H.264 compression format. It is often used as a library in large systems, such as in 
space flight hardware~\footnote{AVN443 encoder: \url{https://www.tvbeurope.com/production-post/visionary-provides-hd-encoders-for-international-space-station}.} where its \emph{(i)} binary size and \emph{(ii)} performance matters. 
Besides, \emph{(iii)} a vulnerability in the \xdeux's (H.264 decoder) function could be used to attack a larger system, such as the case with the Cisco Meeting Servers~\footnote{
A vulnerability: 
\url{https://www.cvedetails.com/cve/CVE-2017-12311/}.}.
Nowadays, the security of modern software systems is mostly threatened internally, that is, by reusing their existing code, without the need for code injection~\cite{sharif2018trimmer}.
This kind of attack allows an attacker to execute arbitrary code on a victim's machine. 
In this attack, the attacker is able to chain some small code sequences (called \textit{gadgets}) and threaten the security of the system. Basically, the exploited code sequences by the attacker end in a return (RET or JMP) instruction. 
Therefore, one of the commonly used metrics for measuring the attack surface in a system is \textit{the number of code reuse gadgets} that are available and which can be exploited by an attacker~\cite{brown2019less,Kurmus:2011:ASR:1972551.1972557,sharif2018trimmer}. 

Motivated by similar requirements, we analyse the effects of 
specializing \xdeux on these three non-functional properties, namely on its binary size, attack surface, and performance. 

\section{Specialization approach}

\revThree{The paper could describe more clearly (e.g., as a table/figure?) the effect of the preprocessor directives. For example: OPTION\_YES 1, OPTION\_NO 1: run-time option not set during compile-time}

\subsection{The vision for debloating unused run-time options at compile-time}
\label{vision}
At the system level, the usual approach is to keep as much variability as possible. All run-time options may somehow be needed one day. Furthermore, packages, binaries, or build instructions force users to take them all. However, this variability is sometimes unnecessary: in a given context, some options may never be tuned and thus they always have the same value for all actual use cases of a given instance of the software. For instance, \mbox{-}\mbox{-}\texttt{cabac} might always have the \emph{true} value. 
In this case, its corresponding variable in the source code becomes a constant. 
By setting the option of \mbox{-}\mbox{-}\texttt{no}\mbox{-}\texttt{cabac} will speed up the video encoding time, but the video quality may deteriorate. Hence, the tune of run-time options depends on the objectives and constraints of the user. 
Keeping an option while actually always using the same value leads to missed opportunities.
First, by knowing that an option has the same value, compilers do not have to make some assumptions and the generated code can be improved. Specifically, compilers can further apply optimizations and 
constant folding, including the propagation of constants and the removal of dead code. 
Secondly, the presence of unused options at run-time is likely 
\textit{(i)} to increase the executable binary size, since the related code will be included; 
\textit{(ii)} can increase the attack surface (\eg the number of code-reuse gadgets~\cite{Kurmus:2011:ASR:1972551.1972557}); and
\textit{(iii)} even slow down the system. 
The vision is that the proactive removal of the unused run-time options can be beneficial to several non-functional properties of software. 

To technically support this vision, developers should have the means to debloat run-time options, for example, 
by lifting them at compile-time, 
and thus specialize the original program. 
With configurable specialization, developers can build specialized packages (binaries) out of the specialized program. For example, one can envision delivering new variants of \xdeux: \emph{\texttt{x264-fast}} -- with fast encoding time, \emph{\texttt{x264-hq}} -- with high quality video encoding, \emph{\texttt{x264-secured}} -- with fewer code-reuse gadgets, or \emph{\texttt{x264-tinyfast}} -- a small binary size yet fast \xdeux.
Importantly, they do not necessarily specify a full and complete configuration: not all options are subject to specialization since a part of the variability can be preserved at run-time for addressing several usages. But, some of the unneeded options that may improve the non-functional properties and the performance of the system can be eliminated from code. 
Our idea is inspired by the multi-staged configuration~\cite{czarnecki2004,hubaux2011} for a step-wise specialization of variability models. In our case, specialization occurs at the code level and regarding the run-time options. 

\subsection{Process for program specialization} 
\label{appraoch}

\begin{figure*}[t]
\includegraphics[width=0.9\textwidth]{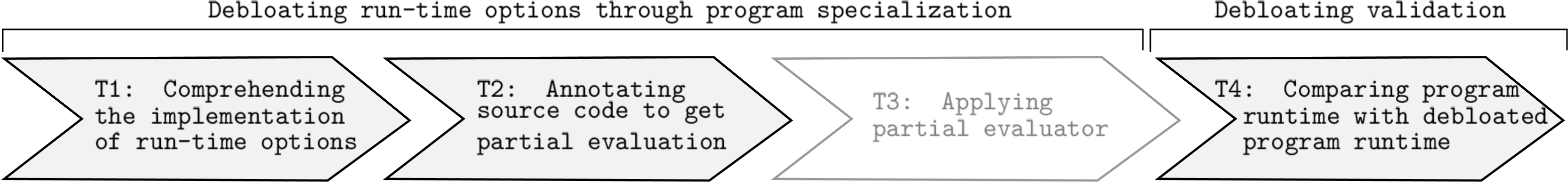}
\caption{Overall process for debloating run-time options through program specialization}
\label{fig:appraoch}
\end{figure*}

To quantify the benefits of such a vision, we report in this paper about our experiments in debloating run-time options of \xdeux through program specialization (\cf~Figure~\ref{fig:appraoch}). To achieve our vision, one should be able to specialize \xdeux by removing run-time configuration options on demand. This implies removing the parts of source code that implement such options, 
which is different from simply disabling those options in a configurator or at the command-line parameter level.
Moreover, to be able to remove the unused run-time options, they first need to be located in the source code.
\revOne{I did not understand the difference between T1 and T2.}
Hence, we propose a removal approach which is based on 4 tasks: (\texttt{T1}) \emph{we first comprehend the implementation of run-time options},
then (\texttt{T2}) \emph{we annotate the corresponding code which implement a given run-time option}, and finally (\texttt{T3}) \emph{we apply a partial evaluator} taking care of the actual code removal. In addition to these three tasks, we conducted the last task (\texttt{T4}) which consists in \emph{validating the program specialization} by comparing the resulting debloated program with the original one according to a given configuration of run-time options.

The experiments conducted for these 4 tasks 
in the context of run-time options in C-based systems, taking \xdeux as an illustrative example, are presented in the following. As a partial evaluator, we decided 
to use the preprocessor switches~\cite{baxter2001preprocessor}
and all the optimizations (\eg dead code elimination) provided by GCC (in grey in~Figure~\ref{fig:appraoch}).
A partial evaluator is a method for simplifying programs when program parameters are known~\cite{jones1993partial,baxter2001preprocessor}. For instance, when it is known that a run-time configuration option in a given system will always be unset.  In this study, the program parameters are the given run-time options for removal whereas their removal and code optimization is completed by the GCC compiler.
Hence, the annotations of the source code consist in adding the required C preprocessor directives to let GCC remove the code corresponding to the given run-time options. Moreover, the validation of the program specialization is not realized at the program level but at the binary level after the complete compilation of the original program.

\paragraph{\texttt{T1}: Comprehending the implementation of run-time options}
\label{mapping}

To debloat unused run-time options later, we first had to comprehend their implementation within the codebase. 
During this process, we identified the following patterns how they are handled in the source code. 
(A)~The \xdeux, and most of the C-based systems, uses the \texttt{getopt.h}~\footnote{\url{https://www.gnu.org/software/libc/manual/html_node/Getopt.html}} which has the functionalities to automate some of the chores involved in parsing typical \texttt{Unix} command-line options.
(B)~A corresponding variable is usually used to represent a run-time option. In this case, we manually analyse the data-flow corresponding to this variable and locate all parts of the code used to implement that particular run-time option.
(C)~Some functions were dedicated to handling a single run-time option. Hence, removing all the functions calls will eventually lead to the removal of the complete functions considered as dead code.
(D)~Then, some of the options have dependencies, for example, \mbox{-}\mbox{-}\texttt{cabac} and \mbox{-}\mbox{-}\texttt{no}\mbox{-}\texttt{cabac} have an alternative logic. This suggests that \xdeux will be unsound if both these options are removed. 
In order to provide inputs for future works about automation (see~Section~\ref{discussions}), these first experiments have been conducted manually to identify the patterns how run-time options are handled within a codebase.

\paragraph{\texttt{T2}: Annotating source code} 
\label{delimiting}

\revOne{Given that evry option is captured in a variable in the C code, would it not be simpler to just replace the variable with a constant and let the compiler do the rest?}
\revTwo{l.97 "partial evaluation": the term is used throughout again and again, but by my understanding you are simply using the preprocessor, and do not actually have an actual partial evaluator applied to the code, beside the standard preprocessor/compiler combination. The problem here is that my expectations are set at the level: "oh, they are going to do something complicated, which might solve some very specific problems", but in the end it's just that a big word is used for relying on the standard compiler. This doesn't seem an ideal way to present this work.}
We now present our approach of removing unused run-time options. 
First, we lift run-time options to compile-time through annotating them with preprocessor directives.
Then, the objective is to let the preprocessor and optimizations of GCC remove the source code related to the values of a given run-time option. 
%
A way to make explicit options' values is to annotate them in places where they are implemented (\ie to locate them).
For this purpose, we propose to use C preprocessor directives, namely, \texttt{\#define}, \texttt{\#if}, and \texttt{\#endif}. They are special instructions directed to the preprocessor, which is part of the compiler, in order to be processed. Because of their prominent and widespread nature, we chose them as they are pretty light to learn, use, and make it possible to remove the enclosed code.

\begin{algorithm}[t]
	\caption{Delimiting a Boolean run-time option} 
	\label{algorithm1}
	\scriptsize
	\begin{algorithmic}[1]
	    \INPUT compilable $\theta$ C-based system with run-time options
	    \OUTPUT compilable $\theta$ with delimited run-time option(s)
	    \CREATE $removeoption.h \in source\_files$ in $\theta$
	    \For {\texttt{option} $\in$ $\theta_{runtime\_options}$}
		\For {\texttt{option} $= true, false$} \hfill  \textcolor{codegreen}{ //  \eg \emph{true} = \mbox{-}\mbox{-}\texttt{cabac},  \emph{false} = \mbox{-}\mbox{-}\texttt{no}\mbox{-}\texttt{cabac}}
		    \State Within $removeoption.h$
		        \State \hskip 0.48cm \texttt{$\#$define OPTION\_YES 1} \hfill \textcolor{codegreen}{ // \eg \texttt{$\#$define CABAC\_YES 1} }
		        \State \hskip 0.48cm \texttt{$\#$define OPTION\_NO 1}  \hfill
		        \textcolor{codegreen}{ // \eg \texttt{$\#$define CABAC\_NO 1} }
		        \State \hskip 0.48cm Set \texttt{OPTION\_YES 0} $\Rightarrow$ $\theta_{specialized} \subset \theta$
		    \State \textbf{repeat}
		    \State \hskip 0.48cm Find code that implement its $true$ value
		    \State \hskip 0.48cm Enclose it within \texttt{$\#$if OPTION\_YES...$\#$endif}
		    \State \textbf{until} \underline{$\theta_{specialized}$ is valid}
		        \State \hskip 0.6cm Set \texttt{OPTION\_YES 1}, 
		        \State \hskip 0.6cm Set \texttt{OPTION\_NO 0} $\Rightarrow$ $\theta_{specialized} \subset \theta$
		    \State \textbf{repeat}
		    \State \hskip 0.48cm Find code that implement its $false$ value
		    \State \hskip 0.48cm Enclose it within \texttt{$\#$if OPTION\_NO...$\#$endif}
		    \State \textbf{until} \underline{$\theta_{specialized}$ is valid}
		        \State \hskip 0.48cm Set \texttt{OPTION\_NO 1}
		\EndFor
	    \EndFor
	\end{algorithmic} 
\end{algorithm}

\begin{figure*}[t]
\centering
\begin{subfigure}{.25\textwidth}
\centering
\begin{lstlisting}[caption={}, language=C]
/* File removeoption.h */
#ifndef CABAC_YES
#define CABAC_YES 0
#endif
#ifndef CABAC_NO
#define CABAC_NO 1
#endif
/* The rest of the directives are omitted */
\end{lstlisting}
\end{subfigure}%
\hfill
\begin{subfigure}{.30\textwidth}
\centering
\begin{lstlisting}[caption={}, language=C]
/* File encoder/encoder.c */
/*The previous code is omitted*/
#if CABAC_YES //option --cabac
    if( h->param.b_cabac )
        x264_cabac_init( h ); 
#endif
#if CABAC_YES && CABAC_NO
    else
#endif
#if CABAC_NO //option --no-cabac
        x264_cavlc_init( h ); 
#endif
/*The rest of the code is omitted*/
\end{lstlisting}
\end{subfigure}
\hfill
\begin{subfigure}{.37\textwidth}
\centering
\begin{lstlisting}[ caption={}, language=C]
/* File common/cabac.c */
/*The previous code is omitted*/
#if CABAC_YES
void x264_cabac_init( x264_t *h ) {
    int ctx_count = CHROMA444 ? 1024 : 460;
    for( int i = 0; i < 4; i++ ) {
    // 8 lines of computation
    }
}
#endif
/*The rest of the code is omitted*/
\end{lstlisting}
\end{subfigure}
\caption{An excerpt from the \texttt{removeoption.h} file with 2 defined preprocessor directives (left). The \texttt{encoder.c} file where \mbox{-}\mbox{-}\texttt{cabac} and \mbox{-}\mbox{-}\texttt{no}\mbox{-}\texttt{cabac} are implemented (middle). The \texttt{cabac.c} file where the directive is used at the function level (right)}
\label{lst:removeoptions}
\end{figure*}

\revOne{I struggled with a lot of the notation in Algorithm 1. What does "Set OPTION\_YES 1" mean, especially when a bit later we have "Set OPTION\_YES 1"? What does Line 6 mean?}

\emph{Algorithm~\ref{algorithm1}} shows the manual approach that we conducted to delimit Boolean run-time options in \xdeux, one at a time, by annotating their \texttt{true} and \texttt{false} values.
First, we make sure that the system is compilable. A \texttt{removeoption.h} file is then added. It contains the defined preprocessor directives, which are used to annotate and remove the desired run-time options.
Hence, for each run-time option that is going to be removed (lines $1-2$, $18-19$), we add a new preprocessor directive (in lines $4-5$ are added two, \eg for \mbox{-}\mbox{-}\texttt{cabac} and \mbox{-}\mbox{-}\texttt{no}\mbox{-}\texttt{cabac}), which directives can take the value $1$ or $0$. Then, we 
set the directive of the first option to $0$ (line $6$)
and 
annotate its respective code (lines $7-10$). To make sure that all parts of code that are relevant to an option are annotated, we check the validity of the given software system after removing each option (lines $10$,$16$).
As soon as the system is valid (\ie we annotate all parts of code that implement the first option, and only those parts), we set the value to $1$ (line $11$). Then, in lines $12-17$ we repeat the same process, but now for the second option. %
%
The algorithm is repeated for as many Boolean options as it is necessary and it can be extended for not Boolean options. 
The process ends as soon as the system after the option's removal is valid.

\emph{Illustrative example.}
To illustrate~Algorithm~\ref{algorithm1}, we provide a concrete example from the \xdeux system.
%
\xdeux has more than one hundred run-time options, one of which is the important Boolean option of \mbox{-}\mbox{-}\texttt{cabac}. In \xdeux, both values of a Boolean run-time option use different names, such as \mbox{-}\mbox{-}\texttt{cabac} and \mbox{-}\mbox{-}\texttt{no}\mbox{-}\texttt{cabac}, therefore we will refer to them as run-time options instead of values of an abstract option. 
In~Figure~\ref{lst:removeoptions} (middle) is shown a code snippet that implements these options.
It has a conditional \texttt{if} statement with two branches where both options \mbox{-}\mbox{-}\texttt{cabac} and \mbox{-}\mbox{-}\texttt{no}\mbox{-}\texttt{cabac} are involved. When the condition is \texttt{true} (lines $4-5$), that is, the user chose to encode a video with the option of \mbox{-}\mbox{-}\texttt{cabac}, then the function on the second branch for \mbox{-}\mbox{-}\texttt{no}\mbox{-}\texttt{cabac} (line $11$) becomes dead code and thus can be removed. Further, a compiler can also realize that some functions inside it are never called and will further optimize the code.

\revTwo{When reaching Fig. 2, I was wondering why the if statement on line 4 hasn't been
considered. While I don't assume that this specific statement has any
performance impact, it's unclear to me that the way the preprocessor has been
used is really going to extract all performance benefits possible. For this
branch for instance, I'd assume that the true branch is always taken when
CABAC\_YES=1 and CABAC\_NO=0.}

In order to locate and remove the related code of an unused option, as it is \mbox{-}\mbox{-}\texttt{no}\mbox{-}\texttt{cabac}, we proposed~Algorithm~\ref{algorithm1}.
Specifically, we first make sure that the system of \xdeux is compilable. We then add the \texttt{removeoption.h} file and two directives, namely \texttt{CABAC\_YES} and \texttt{CABAC\_NO}, as shown in~Listing~\ref{lst:removeoptions} (left). Further, each of them are used to annotate all lines of code related to options \mbox{-}\mbox{-}\texttt{cabac} and \mbox{-}\mbox{-}\texttt{no}\mbox{-}\texttt{cabac}, respectively, as is given in~Listing~\ref{lst:removeoptions} (middle) in lines $3$, $6-7$, $9-10$, and $12$. 
We used preprocessor directives at different granularity levels and in different combinations. In~Listing~\ref{lst:removeoptions} (middle) and (right) is shown their usage to annotate the related code of \mbox{-}\mbox{-}\texttt{cabac} and \mbox{-}\mbox{-}\texttt{no}\mbox{-}\texttt{cabac} at statement and function levels, respectively. 
In this way, we continue to annotate statically the related parts of code for one option at a time until the system of \xdeux becomes again compilable and valid. We check the validity of the program when an option is removed, that is, of its specialized version.
\revFour{The decision to add preprocessor annotations for conditional compilation but to not modify the code any further [line 463] should be justified and discussed better. In particular, it seems that ironing out some leftover conditional statements that now refer to a constant value, could be a good source for optimisation, and it can be done automatically by partially evaluation `if` conditions.}
It should be emphasized that, while we add the run-time options' annotations, we kept unmodified the existing code of the system. For this reason, we add two directives in combination as in lines $7$ and $9$ in~Listing~\ref{lst:removeoptions} (middle). This way of annotating allows the user to remove one of the options by setting the values $(0,1)$ or $(1,0)$  in~Listing~\ref{lst:removeoptions} (left) for each directive respectively. Setting both directives to $0$ is unacceptable as at least one of the options must be part of the system. Whereas, when both directives are set to $1$ then both options are kept available at run-time.   

The part of the approach to ensure system validity and doing its automated specialization are given in the following.

\paragraph{\texttt{T3 \& T4}: Debloating validation}
\label{systemsoundness}

The added preprocessor directives make it possible to remove an unused option in a given context, and thus to specialize the entire given program. But, when annotating an option, we have to make sure that all relevant parts in the source code of a given option are annotated, otherwise removing only some of its lines of code will make the system uncompilable. Then, being only compilable is not enough. The program must also be valid, meaning that the remained options in the program should be intact and the system delivers the same functionality as before when the option was part of the code but unset/disabled. Hence, we check program validity during the annotation of each option. In particular, we check whether the program after removing a given option (\eg \mbox{-}\mbox{-}\texttt{cabac}) (A)~is compilable, (B)~preserves the rest of its behavior, and (C)~is interoperable, that is, when a user tries to use the removed option then the system will provide a pertinent warning message. 

The whole validity check in \xdeux is automated. First, the original 
and specialized program are used to encode 
eight carefully chosen videos (\cf~Section~\ref{availability}) by the ten built-in presets~\footnote{A preset is a set with run-time configuration options that are set at once.} in \xdeux and record the resulting video sizes (A). 
\revTwo{l.604 "10 presets": here I was wondering whether these are built into x264 or provided by you. It becomes a little more clear later, but confused me here.}
In the next check (B), if the size in bytes of each encoded video by the original and specialized program is exactly the same and if using the removed option (\eg  of \mbox{-}\mbox{-}\texttt{cabac}) at run-time in the specialized program is not possible anymore
we then conclude that the specialized program is valid.
In case that at least one of these two conditions is false, then the program is invalid, meaning that the annotated option (\eg of \mbox{-}\mbox{-}\texttt{cabac}) needs to be further improved.
Regarding the interoperability of the specialized program~(C), for example, trying to encode a video by setting the removed option of \mbox{-}\mbox{-}\texttt{cabac} will show a warning message, which notifies the user that the option of  \mbox{-}\mbox{-}\texttt{cabac} is no longer available.
The validation process is specific to our program of interest, here \xdeux. 
However, the methodology is general and essentially requires to develop a testing procedure (\eg oracle~\cite{bruce2018approximate,barroracle}) capable of comparing two programs (\eg their outputs) -- the original and the specialized ones.  


\section{Experimental design}
\label{experimentaldesign}
We outline the design of our experiment via the hypothesis, system specialization scenarios, and experimental settings.



\subsection{Hypothesis}
\label{hypothesis}

We present three hypotheses each to be supported or refuted by our analysis. The first hypothesis concerns the binary size, the second one the attack surface, and the third one the performance of a software system. 
Our null hypotheses are labeled \hyperref[H01]{$H_{01}$}, \hyperref[H02]{$H_{02}$}, and \hyperref[H03]{$H_{03}$}, respectively.
Their respective alternative hypotheses are labeled \hyperref[HA1]{$H_{A1}$}, \hyperref[HA2]{$H_{A2}$}, and \hyperref[HA3]{$H_{A3}$}.


\begin{description}
	\item[$\boldsymbol{H_{01}:}$] \label{H01}
	The baseline software system and its specialization 
	are not significantly different with respect to their binary size.
	\item[$\boldsymbol{H_{A1}:}$] \label{HA1}
	The specialized software system 
	has smaller binary size than its baseline.
	\item[$\boldsymbol{H_{02}:}$] \label{H02}
	The baseline software system and its specialization 
	are not significantly different with respect to their attack surface (\ie $\#$ code reuse gadgets).
	\item[$\boldsymbol{H_{A2}:}$] \label{HA2}
	The specialized software system 
	has smaller attack surface  (\ie $\#$ code reuse gadgets) than its baseline.
	\item[$\boldsymbol{H_{03}:}$] \label{H03}
	The baseline software system and its specialization 
	do not perform significantly different with respect to their encoding time, bitrate, and frame rate. %
	\item[$\boldsymbol{H_{A3}:}$] \label{HA3}
	The specialized software system performs better than its baseline with respect to their encoding time, bitrate, and frame rate.
\end{description}

To statistically evaluate our hypotheses, we build up to 20 specializations of \xdeux. Then, we measure their binary size, attack surface (\ie $\#$ code reuse gadgets), and performance with respect to the encoding time, bitrate, and frame rate of 8 input videos. 
Finally, 
we do a statistical hypothesis testing by using the Wilcoxon signed-rank test~\cite{wilcoxon1963critical}, which is one of the most used statistical test in software engineering~\cite{dybaa2006systematic}. We reject any of the null hypothesis at a common significance level of $\alpha = 0.05$ and do one sided ("greater"/"less")
testing.







\subsection{Baseline system}
\label{baselinesystem}

The system of \xdeux comes with a default configuration, that is, each of its compile-time and run-time options is either enabled or disabled by default. 
In the case of external libraries, provided as compile-time options, \xdeux has six external libraries which are enabled by default. 
Therefore, for our study, we installed all of them in our used environment.
Aside from this, the other compile-time and run-time options are kept to their default values. 
We refer to such an \xdeux's version with default configuration as the \textit{baseline system}.
Put it differently, the baseline system has the current configuration that it has in its respective git repository (\cf its commit,~Section~\ref{casestudy}).
In the following, we will refer to its measured properties as the \textit{baseline binary size}, \textit{baseline number of gadgets}, \textit{baseline encoding time}, \textit{baseline bitrate}, and \textit{baseline frame rate}.

\begin{table}[t]
\caption{Run-time options and presets to specialize \xdeux}
\centering
\label{tbl:bloat}
\scriptsize
\begin{tabular}{@{}p{2cm}
p{0.3cm}p{0.3cm}p{0.3cm}p{0.3cm}p{0.3cm}p{0.3cm}p{0.3cm}p{0.3cm}p{0.3cm}p{0.3cm}@{}}
\toprule
\diagbox{Options}{Presets} 
& \rot{ultrafast} 
& \rot{superfast}
& \rot{veryfast} 
& \rot{faster} 
& \rot{fast} 
& \rot{medium} 
& \rot{slow} 
& \rot{slower} 
& \rot{veryslow} 
& \rot{placebo}\\ 
\multicolumn{1}{l|}{}
& \hl{$S_{11}$} 
& \hl{$S_{12}$}
& \hl{$S_{13}$}
& \hl{$S_{14}$} 
& \hl{$S_{15}$} 
& \hl{$S_{16}$} 
& \hl{$S_{17}$}
& \hl{$S_{18}$}
& \hl{$S_{19}$}
& \hl{$S_{20}$} \\
\midrule
\multicolumn{1}{l|}{\mbox{-}\mbox{-}no\mbox{-}mixed\mbox{-}refs \hl{$S_1$}}
& $\bullet$ 
& $\bullet$ 
& $\bullet$ 
& $\bullet$ 
& $\circ$ 
& $\circ$ 
& $\circ$ 
& $\circ$ 
& $\circ$ 
& $\circ$ \\
\multicolumn{1}{l|}{\mbox{-}\mbox{-}no\mbox{-}mbtree \hskip 0.37cm \hl{$S_2$}} 
& $\bullet$ 
& $\bullet$ 
& $\circ$ 
& $\circ$ 
& $\circ$ 
& $\circ$ 
& $\circ$ 
& $\circ$ 
& $\circ$ 
& $\circ$ \\
\multicolumn{1}{l|}{\mbox{-}\mbox{-}no\mbox{-}cabac \hskip 0.52cm \hl{$S_3$}}
& $\bullet$ 
& $\circ$ 
& $\circ$ 
& $\circ$ 
& $\circ$ 
& $\circ$ 
& $\circ$ 
& $\circ$ 
& $\circ$ 
& $\circ$ \\
\multicolumn{1}{l|}{\mbox{-}\mbox{-}no\mbox{-}weightb \hskip 0.27cm \hl{$S_4$}}
& $\bullet$ 
& $\circ$ 
& $\circ$ 
& $\circ$ 
& $\circ$ 
& $\circ$ 
& $\circ$ 
& $\circ$ 
& $\circ$ 
& $\circ$ \\
\multicolumn{1}{l|}{\mbox{-}\mbox{-}no\mbox{-}psy \hskip 0.73cm \hl{$S_5$}} 
& $\circ$ 
& $\circ$ 
& $\circ$ 
& $\circ$ 
& $\circ$ 
& $\circ$ 
& $\circ$ 
& $\circ$ 
& $\circ$ 
& $\circ$ \\ 
\multicolumn{1}{l|}{\mbox{-}\mbox{-}mixed\mbox{-}refs \hskip 0.35cm \hl{$S_6$}}
& $\circ$ 
& $\circ$ 
& $\circ$ 
& $\circ$ 
& $\bullet$ 
& $\bullet$ 
& $\bullet$ 
& $\bullet$ 
& $\bullet$ 
& $\bullet$ \\
\multicolumn{1}{l|}{\mbox{-}\mbox{-}mbtree \hskip 0.71cm \hl{$S_7$}}
& $\circ$ 
& $\circ$ 
& $\bullet$ 
& $\bullet$ 
& $\bullet$ 
& $\bullet$ 
& $\bullet$ 
& $\bullet$ 
& $\bullet$ 
& $\bullet$ \\
\multicolumn{1}{l|}{\mbox{-}\mbox{-}cabac \hskip 0.86cm \hl{$S_8$}}
& $\circ$ 
& $\bullet$ 
& $\bullet$ 
& $\bullet$ 
& $\bullet$ 
& $\bullet$ 
& $\bullet$ 
& $\bullet$ 
& $\bullet$ 
& $\bullet$ \\
\multicolumn{1}{l|}{\mbox{-}\mbox{-}weightb \hskip 0.6cm \hl{$S_9$}}
& $\circ$ 
& $\bullet$ 
& $\bullet$ 
& $\bullet$ 
& $\bullet$ 
& $\bullet$ 
& $\bullet$ 
& $\bullet$ 
& $\bullet$ 
& $\bullet$ \\
\multicolumn{1}{l|}{\mbox{-}\mbox{-}psy \hskip 0.98cm \hl{$S_{10}$}}
& $\bullet$ 
& $\bullet$ 
& $\bullet$ 
& $\bullet$ 
& $\bullet$ 
& $\bullet$ 
& $\bullet$ 
& $\bullet$ 
& $\bullet$ 
& $\bullet$ \\ 
\bottomrule
\end{tabular}
\end{table}

\subsection{System specialization}
\label{scenarios}
%
The following are the sample of used run-time options to specialize \xdeux and its two specialization scenarios.

\paragraph{The sample of run-time options.}
As given in~Section~\ref{casestudy}, \xdeux has $162$ run-time configuration options and 10 built-in presets. 
A preset is a group of run-time options that are commonly set to quickly encode a video. 
In \xdeux, only $22$ options are used to build its $10$ presets. 
For this exploratory study, we chose a sample of $10$ run-time options in \xdeux that are mainly part of these presets and may help thus to address our hypothesis. 
In~Table~\ref{tbl:bloat} are given the $10$ available presets in \xdeux and the presence of the $10$ chosen run-time options into them.
In particular, we chose $9$ options that are part of at least one preset, $1$ option that is not a part of any preset (\mbox{-}\mbox{-}\texttt{no}\mbox{-}\texttt{psy}), and $1$ option that is not part of the presets (\mbox{-}\mbox{-}\texttt{psy}).  
For example, the preset \texttt{ultrafast} can be used as in the following and it includes up to $17$ options. 
\begin{lstlisting}[label=listingC, caption={}, language=C, numbers=none]
$ x264 --preset=ultrafast -o <output_video> <input_video>
\end{lstlisting}
We have chosen to analyze $5$ of its options, which will be set at once by this preset and hence marked with filled circles ($\bullet$) in~Table~\ref{tbl:bloat}.
The $5$ other chosen options, namely
\mbox{-}\mbox{-}\texttt{mixed}\mbox{-}\texttt{refs},
\mbox{-}\mbox{-}\texttt{cabac},
\mbox{-}\mbox{-}\texttt{mbtree},
\mbox{-}\mbox{-}\texttt{weightb}, and
\mbox{-}\mbox{-}\texttt{no}\mbox{-}\texttt{psy}, we have chosen to specialize the system of \xdeux regarding this preset.
These options are marked with unfilled circles ($\circ$) in~Table~\ref{tbl:bloat}, meaning that they will remain unset in the case when users use \xdeux to always encode videos with the \texttt{ultrafast} preset. 
In the companion page we provide a table with details regarding the set and unset options' directives in each specialization.


\paragraph{Independent and dependent variables.}
To conduct our experiment, we have as control variables: the compile-time options in the subject system, the number and kind of used videos, and the used environment.
Then, we can identify the independent variable: the remained run-time options in \xdeux after its specialization.
The dependent variables for the study were the measures of binary size, number of code reuse gadgets, encoding time, bitrate, and frame rate.

\paragraph{Specialization scenarios.}
By using the 10 chosen run-time options, we follow two scenarios for specializing \xdeux.

\revOne{How were the 20 specialisations chosen?}

\paragraph{$Scenario_1$: \label{SC1}} \textbf{\xdeux is specialized by removing one run-time option at a time.} With the $10$ considered options, there are $10$ possible specializations of \xdeux. They consist of the specializations regarding one of the 10 run-time options (\ie horizontal aspect) given in the first column  
of~Table~\ref{tbl:bloat}. 
\paragraph{$Scenario_2$:  \label{SC2}} \textbf{\xdeux is specialized by removing one preset at a time.} In this case, we specialize \xdeux regarding the $10$ presets by removing the set of unused options within each specific preset. Based on~Table~\ref{tbl:bloat}, they consist of the vertical specializations.
For instance, towards an \texttt{x264-fast} version, \xdeux will be specialized regarding the \texttt{fast} preset by removing its $5$ unused options, which are marked with "$\circ$". 
Hence, in this scenario, there are $10$ other possible specializations of \xdeux. But, they can be grouped into four specializations, for example, \texttt{veryfast} and \texttt{faster} have the same used and unused options and thus both specializations look the same. Despite this, we still analyze them separately as their performance may differ because of their other (un)set options that are different and are kept out of the study. For instance, \texttt{ultrafast} has $17$ options, but only $10$ of them are analyzed.

In total, there is the baseline system of \xdeux and its 20 possible specializations with \hyperref[SC1]{$Scenario_1$} and \hyperref[SC2]{$Scenario_2$}. 
In the following, we will refer to them as $S_0$, for the baseline, and $S_1$ --$S_{20}$, for the specialized systems.

\begin{table}[t]
\caption{The properties of the eight used videos}
\label{tbl:videos}
\small
\begin{tabular}{@{}lrr|lrr@{}}
\toprule
Name  &  Size  & Length  & Name &  Size & Length \\ 
$[.mkv]$ & $[MB]$ & $[sec.]$ & $[.mkv]$ & $[MB]$ & $[sec.]$ \\ \midrule
V\_1\_720x480 & 108.4 & 13 & V\_5\_640x360 & 172.8 & 20 \\
V\_2\_480x360 & 155.6 & 20 & V\_6\_640x360 & 41.5  & 20 \\
V\_3\_640x360 & 165.5 & 19 & V\_7\_640x360 & 207.4 & 20 \\
V\_4\_640x360 & 172.5 & 19 & V\_8\_624x464 & 217.2 & 20 \\
\bottomrule
\end{tabular}
\end{table}

\subsection{Experiment settings} 
\label{availability}
We conducted our experiment on a Linux workstation running Fedora 33 with Intel Core i7-10610U CPU and 15.3~GiB of memory. In all cases, we use the gcc $10.3.1$ compiler to build the baseline and specialized systems of \xdeux.
Some of the done measurements (\cf~\hyperref[H03]{$H_{03}$}) require an input, namely a video. 
Instead of a random selection, we carefully chose $8$ videos based on a recent work by Lesoil \etal~\cite{lesoil2021interaction2}, which properties are given in~Table~\ref{tbl:videos}.
These 8 videos have approximately similar size, resolution, and length, therefore they can be considered as a constant but they are still different videos. Actually, these video 
are evaluated to be as representative enough of 1,300 input videos in the YT UCG dataset~\footnote{UCG dataset: \url{https://media.withyoutube.com/}.}~\cite{lesoil2021interaction2}.
\revTwo{For the evaluation, why were 8 videos chosen? You seem to be subsetting a
larger corpus, which one might consider cheery picking. It would be good to
motivate the number. Why not 10? And how did you decide whether they are
representative? This part of the methodology seems under-explained.}
\revTwo{For the evaluation I was also wondering whether link-time optimization is used,
and whether the compilation is deterministic, or whether you might be measuring
some type of non-determinism including but not limited to effects from parallel
compilation, link-order differences etc. Some of the observed difference could
be explainable by such effects.}
Besides, considering the suggestions by \textit{DB. Stewart}~\cite{stewart2001measuring}, the  encoding time of videos is measured by using the \textit{time} method, and the measurements are repeated 5 times.
\revTwo{Similarly, l.964 has me worried that the benchmarks are not deterministic. I do
not know whether there's any use of non-deterministic input sources in x264, but if
frame rates differ by such amounts, it may call the measured numbers into
question. Could you please clarify what is going on here? It would be also good
to know how the benchmarking is done exactly. You mention that you measure 5
times. Though, this is only explained in sec. 8 instead of where it should be,
in the methodology section. It would also be important to know whether you
recompile the binaries in between to avoid measuring random code placement
changes? How do you avoid measuring turbo boost effects, thermal CPU budgets,
and Address Space Layout Randomization effects?}
To prevent side effects, all experiments are run sequentially and as the only process in the workstation. 


\begin{figure*}[ht]
\centering
\begin{subfigure}{.48\textwidth}
  \centering
  \includegraphics[width=.82\linewidth]{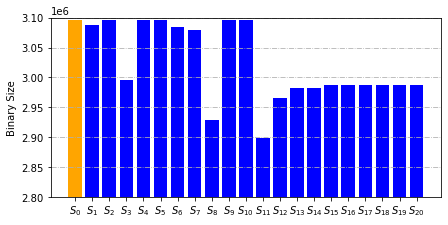}
\end{subfigure}
\begin{subfigure}{.48\textwidth}
  \centering
  \includegraphics[width=.82\linewidth]{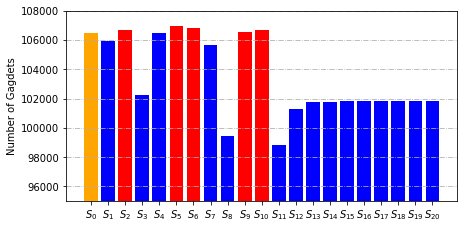}
\end{subfigure}
\caption{The binary size and the 
found number of gadgets in $S_0$ (baseline) and $S_1 - S_{20}$ (specializations).
$(mean \pm std)$ for binary size - $S_0: (3,096,176 \pm 0)$, $S_{1-20}: (3,020,417 \pm 63,641)$, and for gadgets - $S_0: (106,495 \pm 0)$, $S_{1-20}: (103,414 \pm 2,696)$
}
\label{fig:binarysize}
\end{figure*}

\section{Results}
\label{results}

In this section, we present the obtained results. 

\revThree{The overall results of the study (i.e., the tool is usually smaller and faster) are no surprise and the exact numbers are too specific to the example to be generalizable. Nonetheless, the paper contains three larger tables with quantitative results. It would be valuable to describe in more detail how the study itself can be used for further research (e.g., as basis for engineering metrics about run-time vs. compile-time variability?)}

\begin{sidewaystable}
\scriptsize
\caption{The average and standard deviation of 
\textit{encoding time} in seconds [s] of $8$ videos by the baseline ($S_0$) and specialized systems ($S_1-S_{20}$)}
\label{tbl:encodingtime}
\begin{tabular}{p{1.5cm}l|rrrrrrrrrr|r|}
\toprule
System 
& Unit
& ultrafast  
& superfast    
& veryfast    
& faster   
& fast    
& medium  
& slow   
& slower   
& veryslow   
& placebo 
& $H_{03}$\\   \midrule
$S_0$ 
& [s]
& $0.28 \pm 0.02$
& $0.48 \pm 0.01$ 
& $0.74 \pm 0.03$ 
& $1.03 \pm 0.04$ 
& $1.41 \pm 0.04$ 
& $1.73 \pm 0.04$ 
& $2.76 \pm 0.08$ 
& $4.97 \pm 0.16$ 
& $9.85 \pm 0.37$ 
& $35.03 \pm 0.72$ 
& \\ \hline
$S_1$  
& [s]
& - 
& - 
& - 
& - 
& $1.39 \pm 0.03$ 
& $1.77 \pm 0.05$ 
& $2.85 \pm 0.10$  
& $5.08 \pm 0.18$
& $9.86 \pm 0.26$ 
& $35.32 \pm 0.46$ 
& \\
$S_2$  
& [s]
& - 
& - 
& $0.67 \pm 0.02$ 
& $0.96 \pm 0.01$ 
& $1.29 \pm 0.02$ 
& $1.63 \pm 0.04$ 
& $2.61 \pm 0.06$ 
& $4.78 \pm 0.18$ 
& $9.27 \pm 0.42$ 
& $33.23 \pm 0.69$ 
& \\
$S_3$  
& [s]
& - 
& $0.46 \pm 0.02$ 
& $0.73 \pm 0.02$ 
& $1.02 \pm 0.02$  
& $1.43 \pm 0.03$ 
& $1.79 \pm 0.05$  
& $2.80 \pm 0.06$  
& $5.00 \pm 0.18$ 
& $9.96 \pm 0.26$ 
& $35.34 \pm 0.76$ 
& \\
$S_4$  
& [s]
& -  
& $0.46 \pm 0.01$ 
& $0.73 \pm 0.01$ 
& $1.03 \pm 0.01$ 
& $1.43 \pm 0.03$ 
& $1.78 \pm 0.05$  
& $2.92 \pm 0.10$  
& $5.07 \pm 0.19$ 
& $10.11 \pm 0.32$ 
& $35.09 \pm 0.48$ 
& \\
$S_5$ 
& [s]
& $0.29 \pm 0.01$
& $0.46 \pm 0.02$ 
& $0.75 \pm 0.01$ 
& $1.06 \pm 0.03$  
& $1.46 \pm 0.04$  
& $1.83 \pm 0.04$ 
& $2.87 \pm 0.08$ 
& $5.05 \pm 0.19$  
& $10.09 \pm 0.37$ 
& $35.20 \pm 0.63$ 
& \\
$S_6$  
& [s]
& $0.30 \pm 0.01$ 
& $0.47 \pm 0.02$ 
& $0.75 \pm 0.02$ 
& $1.05 \pm 0.01$ 
& - 
& - 
& - 
& - 
& - 
& - 
& \\
$S_7$  
& [s]
& $0.28 \pm 0.01$ 
& $0.45 \pm 0.01$ 
& - 
& - 
& - 
& - 
& - 
& - 
& - 
& -
& \\
$S_8$  
& [s]
& $0.28 \pm 0.01$ 
& - 
& - 
& - 
& - 
& - 
& - 
& - 
& - 
& -
& \\
$S_9$  
& [s]
& $0.29 \pm 0.02$ 
& -  
& - 
& - 
& - 
& - 
& - 
& - 
& - 
& - 
& \\
$S_{10}$  
& [s]
& - 
& -  
& - 
& - 
& - 
& - 
& - 
& - 
& - 
& - 
& \\ \midrule
Avr. $S_{1-10}$ 
& [s]
& $0.29 \pm 0.01$ \textcolor{red}{$\uparrow$}  
& $0.46 \pm 0.01$ \textcolor{ao}{$\downarrow$} 
& $0.72 \pm 0.03$ \textcolor{ao}{$\downarrow$} 
& $1.02 \pm 0.04$ \textcolor{ao}{$\downarrow$} 
& $1.40 \pm 0.07$ \textcolor{ao}{$\downarrow$} 
& $1.76 \pm 0.08$ \textcolor{red}{$\uparrow$} 
& $2.81 \pm 0.12$ \textcolor{red}{$\uparrow$} 
& $4.99 \pm 0.13$ \textcolor{red}{$\uparrow$} 
& $9.86 \pm 0.34$ \textcolor{red}{$\uparrow$} 
& $34.84 \pm 0.90$ \textcolor{ao}{$\downarrow$} 
& \cellcolor{red!15}$\neg$R\\ \midrule
$S_{11}$  
& [s]
& $0.29 \pm 0.01$ 
& - 
& - 
& - 
& - 
& - 
& - 
& - 
& - 
& -
& \\
$S_{12}$  
& [s]
& - 
& $0.47 \pm 0.02$ 
& - 
& - 
& - 
& - 
& - 
& - 
& - 
& -
& \\
$S_{13,14}$  
& [s]
& - 
& - 
& $0.66 \pm 0.03$ 
& $0.95 \pm 0.03$ 
& - 
& - 
& - 
& -  
& - 
& - 
& \\
$S_{15-20}$  
& [s]
& -  
& - 
& - 
& - 
& $1.24 \pm 0.04$ 
& $1.53 \pm 0.03$ 
& $2.56 \pm 0.07$ 
& $4.70 \pm 0.16$ 
& $9.21 \pm 0.40$ 
& $33.55 \pm 0.54$  
& \\ \midrule
Avr. $S_{11-20}$ 
& [s]
& $0.29 \pm 0.0$ \textcolor{red}{$\uparrow$} 
& $0.47 \pm 0.0$  \textcolor{ao}{$\downarrow$} 
& $0.66 \pm 0.0$ \textcolor{ao}{$\downarrow$} 
& $0.95 \pm 0.0$ \textcolor{ao}{$\downarrow$} 
& $1.24 \pm 0.0$ \textcolor{ao}{$\downarrow$} 
& $1.53 \pm 0.0$ \textcolor{ao}{$\downarrow$} 
& $2.56 \pm 0.0$ \textcolor{ao}{$\downarrow$} 
& $4.70 \pm 0.0$ \textcolor{ao}{$\downarrow$} 
& $9.21 \pm 0.0$ \textcolor{ao}{$\downarrow$} 
& $33.55 \pm 0.0$ \textcolor{ao}{$\downarrow$} 
& \cellcolor{green!15}R \\ \midrule \midrule 
Avr. $S_{1-20}$ 
& [s]
& $0.29 \pm 0.01$  \textcolor{red}{$\uparrow$} 
& $0.46 \pm 0.01$ \textcolor{ao}{$\downarrow$} 
& $0.71 \pm 0.04$ \textcolor{ao}{$\downarrow$} 
& $1.01 \pm 0.05$ \textcolor{ao}{$\downarrow$} 
& $1.37 \pm 0.09$  \textcolor{ao}{$\downarrow$} 
& $1.72 \pm 0.12$  \textcolor{ao}{$\downarrow$} 
& $2.77 \pm 0.15$  \textcolor{red}{$\uparrow$} 
& $4.95 \pm 0.16$ \textcolor{ao}{$\downarrow$} 
& $9.75 \pm 0.41$ \textcolor{ao}{$\downarrow$} 
& $34.62 \pm 0.96$ \textcolor{ao}{$\downarrow$} 
& \cellcolor{green!15}R \\ 
$\%$ of $S_{1-20}$ 
&
& \cellcolor{red!15}$1.39\%$ 
& \cellcolor{green!15}$-4.13\%$
& $-4.08\%$
& $-2.12\%$
& $-2.76\%$
& $-0.33\%$ 
& $0.31\%$
& $-0.49\%$ 
& $-1.07\%$
& $-1.18\%$ 
& \\
\bottomrule
\end{tabular}
\end{sidewaystable}

\subsection{The binary size of specialized system}
\label{binarysize}
To address \hyperref[H01]{$H_{01}$}, we measured the binary size of \xdeux 
in bytes. 
At first, under the same experiment settings, we measured the baseline binary size of \xdeux. Then, we specialized \xdeux following \hyperref[SC1]{$Scenario_1$} and \hyperref[SC2]{$Scenario_2$}, measured their respective binary size, and compared them with the baseline binary size. 
Finally, we statistically test the \hyperref[H01]{$H_{01}$} null hypothesis. 

The obtained results in~Figure~\ref{fig:binarysize} (left) show that each specialized system has a smaller binary size than the baseline system. 
The smallest binary size have the specialized systems by \hyperref[SC2]{$Scenario_2$} ($S_{11}$ - $S_{20}$), as they are specialized by more than one run-time option, than those by \hyperref[SC1]{$Scenario_1$} ($S_1$ - $S_{10}$). In percentage, compared to the baseline binary size, the specialized systems $S_1$ - $S_{10}$ have a reduced binary size between $0.001\%$ and $5.416\%$, whereas $S_{11}$ - $S_{20}$ have a reduced binary size between $3.526\%$ and $6.365\%$.
Concretely, the \xdeux's binary size is reduced by $2.45\%$ on average, or up to $6.37\%$ (based on $10$ analysed options). 
To avoid repetition here, all of these data in percentage are also given later in~Table~\ref{tbl:tradeoff}.

Moreover, 
the calculated Wilcoxon signed-rank test shows that the p-value is less than $\alpha = 0.05$ ($t=210$, $p=9.54 \cdot 10^{-7}$)~\footnote{Where $t$ is the sum of the ranks of the differences above or below zero.}, therefore \hyperref[H01]{$H_{01}$} is rejected in favor of \hyperref[HA1]{$H_{A1}$}. This suggests that specializing a software system regarding its run-time options will statistically \emph{significantly} reduce its binary size.
This reduction can be useful for resource-constraint devices. Although there is no direct or strong correlation between binary size and other non-functional properties, the reduction suggests that many paths of the code are eliminated and can be beneficial to the system to be run.

\subsection{The attack surface of specialized system}
\label{gadgets}

To measure the attack surface,
we counted the number of code reuse gadgets in the baseline and $20$ specialized systems of \xdeux. 
For this reason, we used the well-known ROPgadget~\footnote{ROPgadget tool: \url{https://github.com/JonathanSalwan/ROPgadget}} tool and counted the overall gadgets in the system's binary. Then, we statistically test the \hyperref[H02]{$H_{02}$} null hypothesis. 

The resulting overall number of code reuse gadgets in~Figure~\ref{fig:binarysize} (right) show that the overall number of gadgets are most often
reduced (the blue bars) compared to the baseline number of gadgets (the orange bar). 
The specialized systems $S_1$ - $S_{10}$ have fewer gadgets than the baseline system, between $-0.42\%$ and $6.60\%$, whereas specialized systems $S_{11}$ - $S_{20}$ have fewer gadgets, between $4.35\%$ and $7.18\%$. The $-0.42\%$ indicates that some specialized systems by a single option (specifically, the 5 red ones, see also~Table~\ref{tbl:tradeoff}) can have a small increase in the number of gadgets despite that their binary size is reduced. 
This is something that is also claimed by~\cite{brown2019less} that could happen, but not all gadgets in a system can be chained by an attacker and threaten the system. 

However, the calculated Wilcoxon signed-rank test shows that the \hyperref[H02]{$H_{02}$} is not rejected for $S_1$ - $S_{10}$ ($t=35$, $p=0.25$), whereas it is rejected for $S_{11}$ - $S_{20}$ ($t=55$, $p=9.77 \cdot 10^{-4}$). 
In general, it is rejected for $S_{1}$ - $S_{20}$ in favor of the \hyperref[HA2]{$H_{A2}$} ($t=190$, $p=3.54 \cdot 10^{-4}$). 
This suggests that it is better to specialize a software system by multiple run-time options (\ie \hyperref[SC2]{$Scenario_2$}), as this will statistically \emph{significantly} reduce its attack surface (by $2.9\%$ on average, or up to $7.18\%$ based on 10 options in \xdeux), and hence its security will be improved.
Moreover, the attack surface gets reduced much quicker, for $2.9\%$ on average, than the binary size, for $2.5\%$ on average.

\begin{table*}
\scriptsize
\caption{The average and standard deviation of \textit{bitrate} in [kb/s] and \textit{frame rate} in [fps] of $8$ videos by $S_0$ and $S_1-S_{20}$}
\label{tbl:encodingtime02}
\begin{tabular}{ll|rrrrrrrrrr|r|}
\toprule
System 
& Unit 
& ultrafast  
& superfast    
& veryfast    
& faster   
& fast    
& medium  
& slow   
& slower   
& veryslow   
& placebo 
& $H_{03}$ \\   \midrule
$S_0$ 
& [kb/s] 
& $1,800 \pm 0$ 
& $1,167 \pm 0$  
& $905 \pm 0$ 
& $946 \pm 0$ 
& $947 \pm 0$ 
& $935 \pm 0$ 
& $928 \pm 0$ 
& $851 \pm 0$ 
& $843 \pm 0$ 
& $787 \pm 0$
& \\ \midrule
Avr. $S_{1-10}$ 
& [kb/s]  
& $1,800 \pm 0$ \textcolor{blue}{$\leftrightarrow$} 
& $1,167 \pm 0$ \textcolor{blue}{$\leftrightarrow$} 
& $912 \pm 15$  \textcolor{ao}{$\uparrow$} 
& $948 \pm 6$ \textcolor{ao}{$\uparrow$} 
& $946 \pm 3$ \textcolor{red}{$\downarrow$} 
& $935 \pm 2$ \textcolor{red}{$\downarrow$} 
& $926 \pm 4$ \textcolor{red}{$\downarrow$} 
& $849 \pm 4$ \textcolor{red}{$\downarrow$} 
& $849 \pm 14$ \textcolor{ao}{$\uparrow$} 
& $791 \pm 9$ \textcolor{ao}{$\uparrow$} 
& \cellcolor{red!15}$\neg$R\\
Avr. $S_{11-20}$ 
& [kb/s] 
& $1,800 \pm 0$ \textcolor{blue}{$\leftrightarrow$} 
& $1,167 \pm 0$ \textcolor{blue}{$\leftrightarrow$} 
& $938 \pm 0$ \textcolor{ao}{$\uparrow$} 
& $959 \pm 0$ \textcolor{ao}{$\uparrow$} 
& $941 \pm 0$ \textcolor{red}{$\downarrow$} 
& $931 \pm 0$ \textcolor{red}{$\downarrow$} 
& $919 \pm 0$ \textcolor{red}{$\downarrow$} 
& $841 \pm 0$ \textcolor{red}{$\downarrow$} 
& $874 \pm 0$ \textcolor{ao}{$\uparrow$} 
& $806 \pm 0$ \textcolor{ao}{$\uparrow$} 
& \cellcolor{red!15}$\neg$R\\ \midrule
Avr. $S_{1-20}$ 
& [kb/s] 
& $1,800 \pm 0$ \textcolor{blue}{$\leftrightarrow$} 
& $1,167 \pm 0$ \textcolor{blue}{$\leftrightarrow$} 
& $916 \pm 17$ \textcolor{ao}{$\uparrow$} 
& $950 \pm 7$ \textcolor{ao}{$\uparrow$} 
& $945 \pm 3$ \textcolor{red}{$\downarrow$} 
& $934 \pm 2$ \textcolor{red}{$\downarrow$} 
& $925 \pm 5$ \textcolor{red}{$\downarrow$} 
& $848 \pm 5$ \textcolor{red}{$\downarrow$} 
& $853 \pm 16$ \textcolor{ao}{$\uparrow$} 
& $793 \pm 10$ \textcolor{ao}{$\uparrow$} 
& \cellcolor{red!15}$\neg$R \\ 
$\%$ of $S_{1-20}$ 
&  
& $0.00\%$ 
& $0.00\%$ 
& $1.22\%$ 
& $0.46\%$
& $-0.22\%$
& $-0.15\%$ 
& $-0.33\%$ 
& \cellcolor{red!15}$-0.36\%$ 
& \cellcolor{green!15}$1.24\%$ 
& $0.83\%$ 
& \\
\midrule \midrule 
$S_{0}$  
& [fps] 
& $2,257 \pm 157$ 
& $1,313 \pm 43$
& $881 \pm 26$ 
& $590 \pm 13$ 
& $436 \pm 9$ 
& $336 \pm 8$ 
& $218 \pm 6$ 
& $136 \pm 5$ 
& $69 \pm 2$ 
& $18 \pm 1$ 
& \\ \midrule
Avr. $S_{1-10}$ 
& [fps]  
& $2,276 \pm 53$ \textcolor{ao}{$\uparrow$} 
& $1,303 \pm 25$ \textcolor{red}{$\downarrow$} 
& $875 \pm 28$ \textcolor{red}{$\downarrow$} 
& $583 \pm 23$ \textcolor{red}{$\downarrow$} 
& $436 \pm 32$ \textcolor{ao}{$\uparrow$} 
& $337 \pm 24$ \textcolor{ao}{$\uparrow$} 
& $220 \pm 18$ \textcolor{ao}{$\uparrow$} 
& $136 \pm 8$ \textcolor{ao}{$\uparrow$} 
& $69 \pm 4$ \textcolor{red}{$\downarrow$} 
& $18 \pm 1$ \textcolor{ao}{$\uparrow$}
& \cellcolor{red!15}$\neg$R\\
Avr. $S_{11-20}$ 
& [fps] 
& $2,242 \pm 0$ \textcolor{red}{$\downarrow$} 
& $1,301 \pm 0$ \textcolor{red}{$\downarrow$} 
& $957 \pm 0$ \textcolor{ao}{$\uparrow$} 
& $634 \pm 0$ \textcolor{ao}{$\uparrow$} 
& $500 \pm 0$ \textcolor{ao}{$\uparrow$} 
& $396 \pm 0$ \textcolor{ao}{$\uparrow$} 
& $259 \pm 0$ \textcolor{ao}{$\uparrow$} 
& $154 \pm 0$ \textcolor{ao}{$\uparrow$} 
& $77 \pm 0$ \textcolor{ao}{$\uparrow$} 
& $19 \pm 0$ \textcolor{ao}{$\uparrow$} 
& \cellcolor{green!15}R \\ \midrule 
Avr. $S_{1-20}$ 
& [fps] 
& $2,271 \pm 49$ \textcolor{ao}{$\uparrow$} 
& $1,303 \pm 23$ \textcolor{red}{$\downarrow$}  
& $888 \pm 42$ \textcolor{ao}{$\uparrow$} 
& $591 \pm 29$ \textcolor{ao}{$\uparrow$} 
& $446 \pm 39$ \textcolor{ao}{$\uparrow$} 
& $347 \pm 32$ \textcolor{ao}{$\uparrow$} 
& $226 \pm 23$ \textcolor{ao}{$\uparrow$} 
& $139 \pm 10$ \textcolor{ao}{$\uparrow$} 
& $70 \pm 5$ \textcolor{ao}{$\uparrow$} 
& $18 \pm 1$ \textcolor{ao}{$\uparrow$} 
& \cellcolor{green!15}R\\
$\%$ of $S_{1-20}$ 
&  
& $0.59\%$
& \cellcolor{red!15}$-0.73\%$
& $0.83\%$
& $0.30\%$ 
& $2.47\%$ 
& $3.07\%$ 
& \cellcolor{green!15}$3.68\%$ 
& $2.74\%$ 
& $1.95\%$
& $2.84\%$
& \\
\bottomrule
\end{tabular}
\end{table*}


\subsection{The performance of specialized system}
\label{performance}
Next, we examine how a specialized system performs to the end-users. Namely, as \xdeux is a video encoder, we measure the performance of a specialized \xdeux by using three metrics:  
the video \textit{encoding time}, \textit{bitrate}, and \textit{frame rate}.

For this purpose, we encoded $8$ videos (\cf~Table~\ref{tbl:videos}) using the baseline and each of the $20$ specialized systems of \xdeux, but only using their available presets.
For example, $S_6$ is specialized regarding the \mbox{-}\mbox{-}\texttt{mixed}\mbox{-}\texttt{refs} (\cf~Table~\ref{tbl:bloat}). We should encode videos only with $4$ presets in it, namely, with \texttt{ultrafast}, \texttt{superfast}, \texttt{veryfast}, and \texttt{faster}. The rest of the presets should not be used in $S_6$ as they require the option of \mbox{-}\mbox{-}\texttt{mixed}\mbox{-}\texttt{refs}, which is removed from them. In case that they are used, the resulting system's performance is expected to be distorted, and thus unrealistic.
Therefore, only the first $4$ presets (\ie with "$\circ$") are available during the encoding with $S_6$.
As another example, the $S_{20}$ that is specialized regarding the \texttt{placebo} preset should encode videos only with \texttt{fast}, \texttt{medium}, \texttt{slow}, \texttt{slower}, \texttt{veryslow}, and \texttt{placebo} as they share the same used and unused options.
This way of encoding is followed by all $20$ specialized systems.
As for the baseline system, we use all $10$ presets with it.

As for the measured values, we first computed the average of values in $5$ repeated measurements, that is, the average encoding time, bitrate, and frame rate of each video by each available preset in each system. 
Next, we computed the average encoding time, bitrate, and frame rate of all $8$ videos within each preset and system.
Results in~Tables~\ref{tbl:encodingtime} and~\ref{tbl:encodingtime02} show the average values with the standard deviation of specialized systems by the \hyperref[SC1]{$Scenario_1$}, \hyperref[SC2]{$Scenario_2$}, and their overall average, per preset.
The dash ("-") in~Table~\ref{tbl:encodingtime} 
is used to mark the unmeasured values for the unavailable presets in each specialized system. 
It can be noticed that $S_{10}$ has no values, meaning that it has no available presets as all of them require the option of \mbox{-}\mbox{-}\texttt{psy}, which is removed for the specialization of $S_{10}$.
To save space, the results in~Table~\ref{tbl:encodingtime02} are more condensed, showing only these three averages for bitrate and frame rate. 
In percentage, they are more detailed in~Table~\ref{tbl:tradeoff}. 

\paragraph{Encoding time.}
The overall \textit{Average} $S_{1-20}$ in~Table~\ref{tbl:encodingtime} shows that in $8$ from $10$ presets the encoding time is improved or decreased between $0.33\%$ and $4.13\%$, whereas it is increased or worsened in only $2$ presets and for less, for $0.31\%$ and $1.39\%$.
Actually, the increased time in \texttt{ultrafast} and \texttt{slow} is very small, for $3.9$ and $8.4$ milliseconds, respectively. 
Still, as our used system to measure them has a higher resolution (of $1$ microsecond), we believe that the increased  encoding times are a consequence of our x264 specializations.
Looking at the individual specializations, the $S_{15}$ has the most improved encoding time, for $11.97\%$. 
This means that because of its removed unused options the preset \texttt{fast} in $S_{15}$ becomes even faster. Moreover, the encoding time measurements have always less than $0.96$ seconds in terms of standard deviations.

\paragraph{Bitrate.}
As shown in Table~\ref{tbl:encodingtime02}, the number of presets with an equal, increased, or decreased bitrate is the same. Among the specializations, $S_{19}$ has the most increased bitrate, for $3.71\%$, indicating that the video compression will be most significantly improved by this specialization because of its removed unused run-time options. In our measurements, the bitrate standard deviations are minor, less than 17 [kb/s]. 

\paragraph{Frame rate.}
A more notable improvement can be observed in the frame rate in the \textit{Average} $S_{1-20}$, in Table~\ref{tbl:encodingtime02}.
The frame rate is improved in $9$ from the $10$ presets, between $0.30\%$ and $3.68\%$, whereas only in \texttt{superfast} it is worsened, for only $0.73\%$. 
As for the specializations, the $S_{17}$ has the most increased frames per second in videos, for about $18.65\%$, meaning that the video compression will be improved by this specialization. 
The frame rate deviates more in our measurements, up to $157$ [fps], because by default it is auto-detected.

The key observations on the \xdeux's performance are that removing even only $5$ of its unused run-time options will improve the video encoding time in $80\%$ of the presets (for up to $4.13\%$), will improve the bitrate in $40\%$ of the presets (for up to $1.24\%$), and will improve the frame rate in $90\%$ of the presets (for up to $3.86\%$). 
In practical terms, using a specialized system in those predefined configurations, a video is encoded faster and exhibits a better frame rate and bitrate than the original \xdeux.  
Moreover, the calculated Wilcoxon signed-rank test given in the last column in~Tables~\ref{tbl:encodingtime} and~\ref{tbl:encodingtime02} shows that \hyperref[H03]{$H_{03}$} is not rejected for $S_1$ - $S_{10}$ for 
encoding time ($t=28$, $p=0.5$), 
bitrate ($t=10$, $p=0.16$), 
and frame rate ($t=26$, $p=0.46$). 
But, it is rejected for $S_{11}$ - $S_{20}$ for 
encoding time ($t=54$, $p=1.95 \cdot 10^{-3}$) 
and frame rate ($t=7$, $p=0.02$).
In general, it is rejected for all specializations $S_1$ - $S_{20}$ in favor of \hyperref[HA3]{$H_{A3}$} for 
encoding time ($t=51$, $p=0.01$)
and frame rate ($t=7$, $p=0.02$), 
but not for bitrate ($t=10$, $p=0.16$). 
Hence, specializing a software system regarding its unused run-time options could
improve its performance (in \xdeux, it does \emph{not significantly} improve its bitrate, but it statistically \emph{significantly} improves its encoding time and frame rate).

\subsection{The trade-off among the system properties}
\label{tradeoff}
%

\begin{table}[t]
\caption{A comparison of changes, in $\%$, of five properties} 
\label{tbl:tradeoff}
\centering
\scriptsize
\begin{tabular}{@{}lrrrrr@{}}
\toprule
System & Binary size & Gadgets & Encoding time & Bitrate  & Frame rate  \\ \midrule
$S_1$   
& $-0.270\%$  
& $-0.51\%$  
&  $0.89\%$     
&  $0.00\%$ 
&  $-1.78\%$\\
$S_2$   
& \cellcolor{orange!15} $-0.001\%$  
& $0.18\%$  
&  $-5.38\%$    
&  $0.96\%$ 
&  $8.53\%$ \\
$S_3$   
& $-3.254\%$  
&  $-4.00\%$ 
&  $0.87\%$     
&  $0.00\%$ 
&  $-0.65\%$ \\
$S_4$   
& \cellcolor{orange!15} $-0.001\%$  
& $-0.02\%$ 
&  $1.06\%$     
&  $0.00\%$ 
&  $-1.88\%$
\\
$S_5$   
& \cellcolor{orange!15} $-0.001\%$  
&  \cellcolor{red!15} $0.42\%$  
&  $1.29\%$     
&  $0.00\%$ 
&  $-2.26\%$ \\
$S_6$   
& $-0.403\%$  
&  $0.29\%$  
&  $1.07\%$     
&  $0.00\%$ 
&  \cellcolor{red!15} $-2.91\%$ \\
$S_7$   
& $-0.543\%$  
&  $-0.81\%$ 
&  $-4.73\%$    
&  $0.00\%$ 
&  $1.86\%$ \\
$S_8$   
& $-5.416\%$  
&  $-6.60\%$ 
&  $-0.14\%$    
&  $0.00\%$ 
&  $2.32\%$ \\
$S_9$   
& $-0.003\%$  
&  $0.05\%$  
&  $1.58\%$     
&  $0.00\%$ 
&  $2.77\%$ \\
$S_{10}$ 
& \cellcolor{orange!15} $-0,001\%$ 
&  $0.18\%$  
&  NaN          
&  NaN      
&  NaN      \\ \midrule
$S_{11}$ 
& \cellcolor{green!15} $-6,365\%$ 
&  \cellcolor{green!15} $-7.18\%$ 
&  \cellcolor{red!15} $1.71\%$     
&  $0.00\%$ 
&  $-0.68\%$  \\
$S_{12}$ 
& $-4.202\%$ 
&  $-4.88\%$  
&  $-3.07\%$    
&  $0.00\%$ 
&  $-0.91\%$   \\
$S_{13}$ 
& $-3.659\%$ 
&  $-4.43\%$  
&  $-9.58\%$    
&  $2.50\%$ 
&  $8.15\%$  \\
$S_{14}$ 
& $-3.659\%$ 
&  $-4.43\%$  
&  $-9.58\%$    
&  $2.50\%$ 
&  $8.15\%$  \\
$S_{15}$ 
& $-3.526\%$ 
&  $-4.35\%$  
&  \cellcolor{green!15} $-11.97\%$    
&  $-0.67\%$ 
&  $14.77\%$ \\
$S_{16}$ 
& $-3.526\%$ 
&  $-4.35\%$  
&  $-11.28\%$    
&  $-0.45\%$ 
&  $17.64\%$ \\
$S_{17}$ 
& $-3.526\%$ 
&  $-4.35\%$  
&  $-7.24\%$    
&  $-0.98\%$ 
&  \cellcolor{green!15} $18.65\%$ \\
$S_{18}$ 
& $-3.526\%$ 
&  $-4.35\%$  
&  $-5.41\%$    
&  \cellcolor{red!15} $-1.09\%$ 
&  $13.91\%$ \\
$S_{19}$ 
& $-3.526\%$ 
&  $-4.35\%$ 
&  $-6.59\%$    
&  \cellcolor{green!15} $3.71\%$ 
& $12.52\%$ \\
$S_{20}$ 
& $-3.526\%$ 
&  $-4.35\%$  
&  $-4.25\%$    
&  $2.49\%$ 
&  $8.41\%$ \\ \midrule
Avr.     
& $-2.447\%$ 
&  $-2.89\%$   
&  $-3.40\%$    
&  $0.36\%$ 
&  $5.47\%$ \\
\bottomrule
\end{tabular}
\end{table}

The obtained results in~Sections~\ref{binarysize}, \ref{gadgets}, and~\ref{performance}
show that different specialized systems of \xdeux have notably different improvements on two non-functional properties -- namely, on the binary size and attack surface -- and on two performance measurements -- namely, on the video encoding time 
and frame rate.  
Therefore, to understand which removed options most significantly improve the system in all or in most of these five aspects (including the cases with an improvement in the bitrate), we compared the changes of these five properties (from the baseline) for each specialized system. 
Specifically, we compared the changes regarding the binary size and number of gadgets given in~Table~\ref{fig:binarysize}, the encoding time given in~Table~\ref{tbl:encodingtime}, and bitrate with frame rate given in~Table~\ref{tbl:encodingtime02}. The results in percentage are given in~Table~\ref{tbl:tradeoff}.  

It can be observed that the specialized systems $S_{11}-S_{20}$ have far more improved values in each of the five aspects than the $S_{1}-S_{10}$. In fact, the maximum improvements in five aspects are achieved with the specializations $S_{11}-S_{20}$, which percentages are colored in green. The red-colored percentages show that the number of gadgets and frame rate are worsened the most in the set of systems $S_{1}-S_{10}$, whereas the encoding time and bitrate are worsened the most in the set $S_{11}-S_{20}$. 
The percentages in orange show the systems with the smallest improvement on the binary size, as there are only improvements. Whereas the \texttt{NaN} values are the unmeasured values, as all presets use the run-time option \mbox{-}\mbox{-}\texttt{psy} which is used to specialize $S_{10}$, hence measuring them is unrealistic. 
It is interesting to note that specializations that provide an improvement or not are different.
Therefore, one should find a trade-off among them and choose the specialization that best meets their requirements. 
For instance, in resource-constraint devices and in those where security is more important, but the encoding time and frame rate are flexible, then $S_{11}$ is a good choice. But, in case that encoding time matters more than the other properties then $S_{15}$ best fulfills this criterion.
In fact, there are 4 specializations that have an improvement in all five properties - namely, $S_{13}$, $S_{14}$, $S_{19}$, and $S_{20}$. These specializations can be chosen when an improvement is expected on all five properties.
Besides these changes within every single specialized system, the overall average in the last row in~Table~\ref{tbl:tradeoff} reveals that each of the five properties gets improved by specializing the \xdeux system.


The more unused run-time options are removed the greater the benefits gained.
Still, these benefits may vary in the system's binary size, attack surface, and performance, therefore finding their trade-offs to a given usage context is necessary.

\section{Related work}
\label{subsec:related}

We discern the four following areas of related work that are particularly relevant to our investigation. 

\paragraph*{Configurable systems and their non-functional properties}
Software product line engineering and variability management is a well-established research area that led in the past decade to several techniques for supporting \eg orthogonal variability management and derivation of custom variants~\cite{borba2012,apel2013,pohl-etal2005,DBLP:conf/icse/MeinickeWVK20,DBLP:conf/splc/BergerLRGS0CC15}. 
In practice, variability is often moved from compile-time to load-time (\aka encoding variability)~\cite{VONRHEIN2016125}. The consequence is to deploy the whole product line (or configurable system) in the delivered software systems. In our work, we follow the opposite path: we aim to move from run-time to compile-time. We specifically explore the benefits of applying the derivation process at compile-time (instead of at run-time) in order to 
specialize legacy software systems, with a particular focus on non-functional properties. 
Besides, there are numerous works about the non-functional properties and performance of configurable systems (\eg see~\cite{julianasurvey,jamshidi2017transfer,siegmund2015,guo2013,
zhang2015performance,valov2015empirical,temple2016,lesoil2021interaction}).
The goal of this line of research is to study the performance modeling of configurable systems without removing run-time options and without altering the original source code. In contrast, we provide an approach for debloating a configurable system regarding its run-time options. We also show the impact of their removal in the system's non-functional properties. 

\paragraph*{Software debloating}
Furthermore, software debloating has been previously explored, for example, to reduce the size of deployed containers~\cite{10.1145/3106237.3106271}, or the attack surface of specific programs~\cite{217642,sharif2018trimmer,debloatingNIER,xin2020,koo2019,10.1145/3243734.3243838,debloating2021}. 
Often, proposed approaches debloat compiled binaries of a system, or debloat a system by removing its unused libraries or its stale feature toggles. To our best knowledge, none of them debloat software regarding the unused run-time configuration options.
Existing approaches are all inheriting of existing works in program specialization~\cite{10.1145/377769.377778}. 
In~\cite{koo2019}, an approach is proposed to remove feature-specific shared libraries that are needed only when certain configuration directives are
specified by the user in configuration files. In contrast, we target run-time options within the source code. 

\paragraph*{Program slicing}
It is another longstanding method for automatically decomposing programs by analyzing their data flow and control flow~\cite{5010248}. 
Program slicing techniques require the identification of a variable of interest, and the benefits have been demonstrated in the context of white-box activities, such as debugging, testing, maintenance, and understanding of programs. However, as in our study, configuration options are usually given from an external (black-box) view of the program, and it is challenging to relate internal variables to such high-level configuration options. 
This has been demonstrated in~\cite{10.1145/2642937.2643001}.
In practice, we also observed much more complex data and control flows that would limit the applicability of such an approach. Hence we believe that such tools can be helpful to partly \emph{support} the task of annotating run-time options in the source code (\cf~Section~\ref{discussions}).
 
\paragraph*{Performance}
Performance has been investigated at run-time, \eg speculative execution~\cite{Gabbay96speculativeexecution} and ahead-of-time compilation~\cite{proebsting1997toba}. Such speedups come in complement to the proposed specialization
approach applied at compile-time. Since the specialization is applied at the source level, existing approaches for optimizing compilers come also in complement to a previous debloating. 
Performance has been also investigated in the context of approximate computing~\cite{10.1145/2893356}, which provides unsound transformations, still useful in the context of possible trade-offs with accuracy. For example, loop perforation~\cite{10.1145/1806799.1806808} is a useful interpolation technique while unrolling relaxable loops (\eg signal processing). In contrast, we explore a sound, compile-time derivation of a specialized configuration space. 

In general terms and in contrast to the related works, we propose in this paper a unique experimental study that combines the system's \emph{variability management} and \emph{configuration debloating} with a specific focus on its \emph{non-functional properties}, including its \emph{performance}. The program slicing techniques with the identified patterns to implement run-time options (\cf~\textbf{T1}) and program transformations can be helpful in the future to automate the specialization process. 

\section{Discussions}
\label{discussions}

\paragraph{Replicability and confirmatory studies.}
In the companion webpage~\footnote{Companion page: \url{https://anonymous.4open.science/r/x264-6ED7}}, we provide the annotated source code of \xdeux as well as the commits together with comments and related issues that document how we annotated it during a period of 12 weeks. 
All measured data for each specialization of \xdeux are available, including  a dockerfile and a usage guide for reproducing the experiment.
We consider that it is fully possible to reproduce/replicate our results and extend our work with, for example, other metrics.

One of the contributions of this paper is the design of the study and the experimental protocol. The exploratory nature of this study was mainly here to generate hypotheses about the possible benefits of specializing the run-time configuration space of a system. Our results call to further investigate this idea on other software systems: Can we observe similar benefits in another application domains and software systems? Is there another non-functional property for which the specialization pays off? 
What are run-time options worth to be removed? 
Is it always safe for the unused run-time options to be changed at compile-time (their pros and cons),
\eg the system should be re-build to enable a compile-time option?
We believe our study can be replicated and we encourage researchers to do so. 
There are several aspects of the experience that deserve special attention, particularly in view of the lessons learned from conducting this case study. 

\paragraph{Manual vs. automated specialization.}
We did rely on a semi-automated specialization process by manually adding annotations to the source code and then automatically specializing the \xdeux.
The choice of a semi-automated process was on purpose and for the sake of the experiments since our primary goal was to control in a fine-grained way the removal of the code. One lesson learned is that manually tracing run-time options is an error-prone activity. It is easy to forget an annotation or break some functionalities. 
Hence, the added T4 task automatically checks that the remaining configuration space of the specialized system remains equivalent to the configuration space of the original system.

The major benefit of our semi-automated approach is that we have a ground truth. 
It is now possible to envision the use of more automated, perhaps more aggressive, techniques and possibly compare the outcome. 
There are two possible directions. The first approach is to use automated techniques to support developers during the task of manually tracing the options in the source code (corresponding to T1 in~Figure~\ref{fig:appraoch}). Works such as feature location or program slicing can be helpful (\eg see~\cite{10.1145/2642937.2643001,ribeiro2010emergent,DBLP:conf/icsr/AL-MsiedeenSHUVS13,5010248,RubinC13,struber2019facing}),
but annotating already implemented run-time options deserves specific attention. Our experience with \xdeux is that the control flow of parameters' values is hard to follow, run-time options are scattered in different files and functions, and there are patterns of annotation that deserve some formalization to either discipline or automate the effort (corresponding to T2). 
 A second approach is to fully automate the specialization, possibly at the binary level. As discussed earlier, most of the debloating techniques have been designed to entirely specialize a system (\ie on a fixed configuration or use case). Our debloating occurs at the options level and for different use case specializations, hence it is unclear how to adapt existing techniques. 
 The risk of automation is that the removal of the code can be too 
 aggressive (unsound) or miss some opportunities (incomplete). 
 In particular, guaranteeing that the remaining configurations preserve their 
 behavior is challenging (corresponding to T2 and T3). 
 A third 
 challenge consists in validating the specialization process. In our study, we compare the outputs (videos) on some representative configurations of the original system and the debloated system. 
 Other validation and comparison procedures can be employed depending on the specificity of the kinds of outputs and software systems (corresponding to T4).

\paragraph{Why and when specialization pays off.} 
 Results show that specialization has varying effects on performance-like properties. 
There are statistically significant improvements, but we also notice 
one 
insignificant improvement (on bitrate). 
We hypothesize that there is a complex interplay between the compiler, the annotated code to remove, and the run-time usage. 
Understanding the reasons why and when compilers can leverage the new specialized source code is an open issue. 
It can help to prioritize the options worth debloating.

\revOne{The differences in quality properties between the different variants appear very small. Is it really worth the whole effort? This should be discussed in a more differentiated manner.}

\section{Threats to validity}
\label{threats}

\paragraph{External threats.} With respect to generalization, we use only a single software subject in a specific domain (video compression) that is implemented in 
C language. \xdeux has been
studied in several papers on configurable software (\eg~\cite{guo2013,siegmund2015,temple2017a,jamshidi2017transfer,julianasurvey}), but not in the context of specializing configurable systems and debloating run-time options.
Our exploratory research aims to investigate a problem that has not been studied or thoroughly investigated in the past: the potential benefits of removing run-time options.  
As with any exploratory case study, we cannot draw any statistically generalizing conclusions from such studies~\cite{ABCofSE}. However, such generalization of findings is not the goal of such studies -- instead, we aim to develop an understanding and propose hypotheses about other similar configurable systems. 

\paragraph{Internal threats.} The performance, such as the video's encoding time by \xdeux,
may differ depending on the inputs processed. To mitigate this threat, we use 8 videos considered as representatives of 1,300 videos coming from the YouTube UGC dataset. 
Then, the measurements are repeated 5 times. 
A threat to validity is also the set of run-time options we consider. Choosing different options may have a different effect on the performance measurements.
Still, we have selected a diverse set of options with potentially different impacts on non-functional properties and which are part of the presets that are widely used and represent a variety of usage in \xdeux. 

Another threat to our experiments is the annotation of run-time options, which can be incomplete and not valid. 
An incomplete annotated option may impact different properties (\eg binary size) and  break existing functionalities, not related to the removed options. 
To mitigate this threat, we manually and thoroughly annotated the source code. We systematically reported on progress through GitHub issues and concrete meetings for clarifying some subtle cases about \eg the meaning of debloating. 
In terms of validity, we used oracles that control that some remaining configurations produce the exact same videos. 
In terms of completeness, we chose to not consider the assembly code of \xdeux when annotating options. A few lines of code and options are involved, suggesting little to negligible impact on our results.

\revOne{How did you check completeness of annotations? Was there some independent review of annotations?}

\section{Conclusion}
\label{conclusion}

We introduced the idea of specializing in a controlled and configurable way the unused run-time configuration space of a software system. 
With our proposed approach, we synthesize a specialized source code at compile time without the run-time options considered unnecessary while preserving other functionalities and remaining configurations.
Using the well-known video encoder of \xdeux as a subject, we showed how to 
specialize a system by changing
the binding time of its run-time options at compile-time. We experimented with 10 of its options over 8 inputs (videos) and over 10 presets of the system's configurations. The results show that specializing a system regarding its unused run-time options brings statistically significant benefits \wrt three non-functional properties, namely binary size, attack surface, and performance (except for bitrate). For instance, removing the single option of \mbox{-}\mbox{-}\texttt{cabac} in \xdeux reduces the binary size by about 7$\%$, its attack surface by 8$\%$, and the video encoding time by 3$\%$.
The combination of options during the specialization process can lead to further benefits (up to 18\% for frame rate). 
We believe that developers, operationals, and users can use our approach to specify and then remove run-time options that are never used or changed for their specific use cases. For example, run-time options involving specific algorithms that do not meet the requirements of the application represent a "bloat" and have no reason to be embedded.


Our exploratory study naturally calls to replicate the approach and results in other software engineering contexts. 
We also brought a new problem to the community: instead of adding options, the challenge of specialization is to remove code related to options. 
As reported in our experience, this is a change that requires revisiting code analysis and feature location techniques traditionally designed for adding or maintaining options, not necessarily for removing them. 
Another future work direction is also to measure the change of the system's non-functional properties 
when some of its compile-time and run-time options are both removed.

\balance 
\bibliographystyle{ACM-Reference-Format}
\bibliography{acmart}


\end{document}